\documentclass[manuscript,screen, nonacm]{acmart}
\usepackage{array}
\AtBeginDocument{%
  \providecommand\BibTeX{{%
    \normalfont B\kern-0.5em{\scshape i\kern-0.25em b}\kern-0.8em\TeX}}}

\setcopyright{acmcopyright}
\copyrightyear{2023}
\acmYear{2023}
\setcopyright{rightsretained}
\acmConference[AutomotiveUI '23 Adjunct]{15th International Conference on
Automotive User Interfaces and Interactive Vehicular Applications}{September
18--22, 2023}{Ingolstadt, Germany}
\acmBooktitle{15th International Conference on Automotive User Interfaces and
Interactive Vehicular Applications (AutomotiveUI '23 Adjunct), September
18--22, 2023, Ingolstadt, Germany}\acmDOI{10.1145/3581961.3609837}
\acmISBN{979-8-4007-0112-2/23/09}

\begin{document}

\title[Holistic HMI Design for Automated Vehicle]{Holistic HMI Design for Automated Vehicles: Bridging In-Vehicle and External Communication}

\author{Haoyu Dong}
\authornote{Both authors contributed equally to this research.}
\email{h.dong@tue.nl}
\affiliation{%
  \institution{Eindhoven University of Technology}
  \city{Eindhoven}
  \country{Netherlands}
  \postcode{}
}
\orcid{0000-0002-7164-7608}

\author{Tram Thi Minh Tran}
\authornotemark[1]
\email{tram.tran@sydney.edu.au}
\affiliation{Design Lab,%
  \institution{The University of Sydney}
  \streetaddress{}
  \city{Sydney}
  \country{Australia}
  \postcode{}
}
\orcid{0000-0002-4958-2465}

\author{Pavlo Bazilinskyy}
\email{p.bazilinskyy@tue.nl}
\affiliation{%
  \institution{Eindhoven University of Technology}
  \city{Eindhoven}
  \country{Netherlands}
  \postcode{}
}
\orcid{0000-0001-9565-8240}

\author{Marius Hoggenmüller}
\email{marius.hoggenmuller@sydney.edu.au}
\affiliation{Design Lab,%
  \institution{The University of Sydney}
  \streetaddress{}
  \city{Sydney}
  \country{Australia}
  \postcode{}
}
\orcid{0000-0002-8893-5729}

\author{Debargha Dey}
\email{debargha.dey@cornell.edu}
\affiliation{Information Sciences,%
  \institution{Cornell Tech}
  \streetaddress{}
  \city{New York}
  \country{United States}
  \postcode{}
}
\orcid{0000-0001-9266-0126}

\author{Silvia Cazacu}
\email{silvia.cazacu-bucica@kuleuven.be}
\affiliation{%
  \institution{Katholieke Universiteit Leuven}
  \streetaddress{}
  \city{Leuven}
  \country{Belgium}
  \postcode{}
}
\orcid{0000-0002-7952-0919}

\author{Mervyn Franssen}
\email{m.franssen@tue.nl}
\affiliation{%
  \institution{Eindhoven University of Technology}
  \city{Eindhoven}
  \country{Netherlands}
  \postcode{}
}
\orcid{0009-0009-3095-4829}

\author{Ruolin Gao}
\email{r.gao@tue.nl}
\affiliation{%
  \institution{Eindhoven University of Technology}
  \city{Eindhoven}
  \country{Netherlands}
  \postcode{}
}
\orcid{0009-0003-8898-5093}

\renewcommand{\shortauthors}{Dong and Tran, et al.}




\begin{abstract}
As the field of automated vehicles (AVs) advances, it has become increasingly critical to develop human-machine interfaces (HMI) for both internal and external communication. Critical dialogue is emerging around the potential necessity for a holistic approach to HMI designs, which promotes the integration of both in-vehicle user and external road user perspectives. This approach aims to create a unified and coherent experience for different stakeholders interacting with AVs. This workshop seeks to bring together designers, engineers, researchers, and other stakeholders to delve into relevant use cases, exploring the potential advantages and challenges of this approach. The insights generated from this workshop aim to inform further design and research in the development of coherent HMIs for AVs, ultimately for more seamless integration of AVs into existing traffic.
\end{abstract}





\maketitle

\section{Background}

The widespread adoption of automated vehicles (AVs) hinges on effective communication with not only internal users, such as drivers and passengers, but also external road users, including pedestrians, cyclists, and manual car drivers~\cite{Hollander2021, dey2020taming, tabone2021vulnerable, colley2022investigating, avsar2021efficient, papakostopoulos2021effect}. Effective communication is crucial for ensuring safety and seamless integration of AV into the existing traffic ecosystem~\cite{Bengler2020HMIs}. In this context, there has been extensive research regarding the design of human-machine interfaces (HMIs) for AVs, often focusing either on internal interfaces (iHMIs) or external interfaces (eHMIs). 
This reductionist approach simplifies the complex design space, excelling at addressing individual components of the system~\cite{Ackoff1973reductionist}—such as perfecting AV-to-driver or AV-to-pedestrian communications separately. 



While the reductionist approach remains effective, current discourse suggests the need for a complementary holistic approach in HMI design for AVs~\cite{Zimmerman2007RtDforIxD}.
This approach is particularly essential for designing for situations where multiple road users interact simultaneously with the AVs through either iHMIs or eHMIs or even both. By examining these interfaces concurrently, we can better understand their dynamics, thereby fostering a consistent experience for all stakeholders involved in the mixed traffic ecosystem~\cite{Karapanos2013}. 


Existing literature on holistic HMI design for AVs remains sparse, with few related works. For example, \citet{Bengler2020HMIs} proposed a framework that encompasses various HMI types (including dynamic HMI (dHMI), vehicle HMI, infotainment HMI, automation HMI, and external HMI), emphasising the importance of dHMI in communicating simultaneously with in-vehicle users and other road users. 

Building on the concept of dHMI, consider an AV equipped with an interactive lighting system as an example of a holistic HMI design. 
For instance, when the AV is about to make a turn or stop, the external lights would flash in a specific pattern to signal its intention to other road users (e.g., \cite{ford2017}). Simultaneously, an internal display would mirror this pattern, informing the passengers about the AV's upcoming actions (e.g., \cite{wilbrink2020reflecting}). This way ensures consistent information being shared with all involved parties. 

However, the holistic HMI approach goes beyond information mirroring. It also involves providing a coherent design language and user experience that caters to the fluid roles that individuals play in the traffic environment. For instance, an individual may transition from being a pedestrian to a passenger in an AV and later interact with AVs as a manual car driver. Even within these roles, the nature of interaction with the AVs can vary, depending on different perspectives.
For example, drivers may be considered as both internal and external users, depending on whether they are operating their own vehicles or interacting with AVs. Recent studies have explored the use of eHMIs to communicate with conventional drivers~\cite{rettenmaier2020after, lingam2022ehmi, papakostopoulos2021effect} as well as conducted comparative analyses on the use of eHMIs and iHMIs for communication with these drivers~\cite{Li2021, avetisyan2023investigating}. This dual role of drivers suggests the potential implication for a design approach that ensures a high degree of compatibility between eHMIs and iHMIs. 

Given the complexity of roles and the potential benefits of a holistic HMI design approach, we see a timely opportunity for further exploration. As a starting point, we propose this workshop as a part of the AutomotiveUI conference.

\section{Workshop Goals}

The workshop is designed with two primary goals:

\begin{enumerate}

    \item \textit{Exploring relevant use cases:} Recognising that a holistic HMI design approach may not be necessary for every scenario, we will facilitate discussions to identify real-world use cases and scenarios where adopting such an approach could be advantageous.
    \item \textit{Building community:} We will bring together researchers and designers specialising in internal and external automotive HMIs, provoking discussion regarding existing design cases and concepts while encouraging the adoption of a broader perspective.  



\end{enumerate}



\section{Expected Outcomes}
This workshop will serve as a foundational exploration of HMI for AV design and research through a holistic lens, cultivating a shared understanding among participants from various backgrounds. The insights derived from our discussions, particularly regarding use cases and validity arguments for this holistic approach, will be compiled in a summary report. We intend to disseminate these findings through an academic article, marking the first consensus among experts in the field.

\section{Workshop Overview}
Our target participants include designers, engineers, and researchers with expertise in automotive HMI design and development, encompassing both in-vehicle and external interfaces. Additionally, we welcome individuals with other profiles, including urban planners, traffic policymakers, etc., to promote interdisciplinary collaboration. 

In order to effectively support the discussion among participants from different backgrounds, we will conduct a deliberately designed collaborative activity known as \textit{participatory physicalisation}~\cite{opportunities2015jansen, whatisdata2022koningsbruggen, getphysical2017huron}. This activity employs detailed instructions and tangible artefacts to facilitate the exploration of ideas, physicalise the discussion process and lead to shared understandings~\cite{panagiotidou2022datavisualisation}. 

We anticipate hosting approximately 30 participants for a well-rounded and insightful session.


\subsection{Pre-Workshop Plan}
The workshop call will be shared across social media platforms like Twitter and LinkedIn 1.5 months prior to the event. To further promote the workshop, organisers will leverage academic and industry networks and encourage attendees to share details within their professional circles.

Participants will be requested to prepare a visual representation of their research topic, which may manifest in diverse forms, including prototype images, graphs or charts derived from their publication, or conceptual frameworks or models. 
These materials serve a twofold purpose: firstly, to offer a clear depiction of each participant's work, and secondly, to facilitate the creation of physicalisation during the collaborative activity proposed for this workshop. The participants can print the visual representation by themselves, or they could submit the file to the organisers prior to the workshop.





\subsection{Workshop Activities}
The workshop will span over a duration of 3 to 3.5 hours with designed activities to achieve the goals (see Table~\ref{tab:schedule}).


\begin{table}[h]
  \small
  \caption{Tentative schedule activities of the workshop}
  \label{tab:schedule}
  \begin{tabular}{m{20em} | m{5em}}
    \toprule
    \textbf{Activity} &\textbf{Time}\\
    \midrule
    Opening \& Introduction & 10 min \\
    \textbf{Keynote} & 30 min \\
    Coffee break & 15 min \\
    \textbf{Participatory Physicalisation - Part 1} & 45 min\\
    Coffee break & 15 min \\
    \textbf{Participatory Physicalisation - Part 2} & 45 min \\
    Closing & 30 min \\
  \bottomrule
\end{tabular}
\end{table}

\subsubsection{Opening \& Introduction} The workshop will begin with the organisers' introductions and a presentation of the workshop's objectives, schedule of activities and expected outcomes.

\subsubsection{Keynote} We will host two distinct guest speakers—one specialising in iHMIs and the other in eHMIs, to present overviews of the state-of-the-art research in their respective fields. 


\subsubsection{Participatory Physicalisation}

This activity consists of two parts separated by a short coffee break. 

\begin{itemize}

\item \textit{Part 1 - Group Discussions}: Participants will be divided into groups of five and engage in a series of timed discussions. Each group will be guided by one of the organising members. The discussions aim to brainstorm potential real-world scenarios where it is crucial to consider both internal and external user perspectives when designing HMIs. To aid these discussions, we will provide basic prototyping tools such as sticky notes, tapes, pencils, and markers. The printed visual representation will be used in this discussion. In addition, unique tangible items, purposefully crafted for this workshop, will be available to stimulate the brainstorming process. The purpose of the tangible items is to provide a physicalisation of the collaborative discussion in order to aid discourse building and scenario exploration. 

\item \textit{Part 2 - Plenary Presentations}: After the group discussions, all participants will join together and each group will present their discussion summaries using the physicalisation as a reference. In this way, we aim to create connections that go beyond previously created groups, identifying points of alignment and friction that contribute to future research directions in holistic HMI design.
\end{itemize}

\section{Organisers}
\textbf{Haoyu Dong}, corresponding author, is a doctoral candidate at the Eindhoven University of Technology in the Department of Industrial Design. Her research follows Research through Design approach and focuses on interaction design research of in-vehicle users and AVs, specifically in non-critical scenarios. 

\textbf{Tram Thi Minh Tran} is a doctoral candidate in the Urban Interfaces Lab at the University of Sydney's School of Architecture, Design, and Planning. Her research investigates the intricate interactions between pedestrians and AVs in complex traffic scenarios involving multiple road users. 

\textbf{Pavlo Bazilinskyy} is an assistant professor at Eindhoven University of Technology focusing on AI-driven interaction between AVs and other road users. He finished his PhD at TU Delft in auditory feedback for automated driving as a Marie Curie Fellow, where he also worked as a postdoc. He was the head of data research at SD-Insights. Pavlo is a treasurer of the Marie Curie Alumni Association (MCAA).

\textbf{Marius Hoggenmüller} is a Lecturer in Interaction Design in the Design Lab at the University of Sydney's School of Architecture, Design, and Planning. His work focuses on prototyping interactions with emerging technologies in cities, such as urban robots and autonomous systems.

\textbf{Debargha Dey} is a postdoctoral researcher at Cornell Tech, with a research focus on human-automation interaction. He has 8+ years of experience in the domain of human factors for automated driving and traffic behaviour.

\textbf{Silvia Cazacu} is a doctoral candidate at the Spatial Applications Division Leuven of KU Leuven, Belgium. Her research 
focuses on participatory physicalisation as an approach to support collaboration in data ecosystems and draws on critical data studies and data humanism. 

\textbf{Mervyn Franssen} is a doctoral candidate at the Eindhoven University of Technology in the Department of Industrial Design. His research is focused on designing and validating natural interfaces and behavioural characteristics for automation systems in order to reduce uncertainty and avoid unpredictable behaviour for human operators who sometimes need to take control.

\textbf{Ruolin Gao} is a doctoral candidate in the Department of Industrial Design, Eindhoven University of Technology. She focuses on the interaction between AVs and manually-driven vehicles to foster smooth interaction in a mixed-traffic environment.


\bibliographystyle{Example/ACM-Reference-Format}
\bibliography{reference}

\appendix

\end{document}